\documentclass{wccm2014}

\usepackage{graphicx}
\usepackage{dcolumn}
\usepackage{amsmath}
\usepackage{amsfonts}
\usepackage{amssymb}
\usepackage{subfigure}
\usepackage{floatrow}
\usepackage{url}

\newcommand{\degrees}{\ensuremath{^\circ}}

\title{BOUNDARY LAYER ADAPTIVITY FOR INCOMPRESSIBLE TURBULENT FLOWS}

\author{KEDAR C. CHITALE$^{*}$, MICHEL RASQUIN$^{\mp}$, ONKAR SAHNI$^{*}$, MARK S. SHEPHARD$^{*}$ AND KENNETH E. JANSEN$^{\dag}$}

\heading{Kedar C. Chitale, et al.}

\address{$^{*}$ Scientific Computation Research Center (SCOREC)\\
Rensselaer Polytechnic Institute\\
CII 4011, 110 8th Street Troy NY 12180, USA \\
e-mail: chitak2@rpi.edu, web page: http://www.scorec.rpi.edu
\and
$^{\mp}$ Leadership Computing Facility \\
Argonne National Laboratory\\
Argonne IL 60439, USA
\and
$^{\dag}$Department of Aerospace Engineering Sciences\\
University of Colorado Boulder\\
Boulder CO 80302, USA
}
\keywords{Computational Fluid Dynamics, Adaptive Mesh Generation, Unstructured Meshing, Turbulent Flows, Boundary Layers}

\abstract{Boundary layers in turbulent flows require fine grid spacings near the walls which depend on the choice of turbulence model. To satisfy these requirements a semi-structured mesh is generally used in this area with orthogonal and layered elements. Adaptation of such a mesh needs to take into account the flow physics along with the standard error indicator approach. In this paper a novel methodology which combines Hessian based error indicators with flow physics to drive mesh adaptation is illustrated. Particular focus is on the thickness adaptation of the layered mesh. The technique is applied to two turbulent incompressible flow cases and its effectiveness is studied.}

\begin{document}
\section{INTRODUCTION}

The accuracy of numerical simulations strongly depends on the mesh resolution and quality. In complex flow problems, it is difficult to determine the adequate mesh resolution {\it a priori}. In such cases, an initial mesh is used to get an approximate flow solution; this mesh is then adapted using {\it a posteriori} error or correction indicators, i.e., based on the approximate numerical solution. This process is carried out iteratively to attain a given level of accuracy. In order for the overall adaptive process to be efficient, the resolution needs to be changed or adapted in a local fashion. This can be done by locally modifying the mesh elements based on a size field. One option is to use scalar error indicators to determine the desired mesh size field, leading to isotropic elements. However, most flow problems of interest exhibit highly anisotropic solution features such as boundary layers, shear layers, shock waves etc. These features are most efficiently resolved with anisotropic elements, i.e., where elements are oriented and stretched in a certain manner.

For viscous flows, boundary layers (BL) need to be resolved efficiently and accurately as they are a prominent flow feature. Additionally when the boundary layers are turbulent, which is often the case for high Reynolds number flows, mesh spacing needs to be properly controlled. Meshing boundary layer region with isotropic elements will put an excessive demand on the computational resources due to an extremely large mesh. Furthermore, a fully unstructured anisotropic mesh results in poorly shaped elements (e.g., elements with aspect ratio above 10,000) and in-turn leads to a numerical solution of poor quality~\cite{Sahni}. To remedy these problems, layered, orthogonal and graded elements are used near the walls whereas rest of the domain is filled with unstructured elements; this is referred to as a {\it boundary layer mesh}. 
% are used near the no-slip walls and the rest of the flow region is meshed with unstructured elements.
Such hybrid meshes have been extensively used for flow simulations~\cite{Garimella}. During adaptivity, it is desirable to maintain this layered structure of elements. Adaptation procedures based on local mesh modifications for boundary layer meshes have been presented in previous work~\cite{Sahni}. These procedures have recently been extended to work in parallel for distributed boundary layer meshes~\cite{Ovcharenko} and have been applied to complicated aerodynamic geometries~\cite{ChitaleMEW}. One of the main limitations of the adaptive strategies described in the above references is that the wall normal mesh spacing in the boundary layer is kept constant. The next step in the area of adaptive research for boundary layers is the ability to adapt in the wall normal direction and finding suitable indicators to set the mesh sizes in that direction. 
% In this study, we use these procedures for transonic and turbulent flows, where to control the adaptive process we use flow physics indicators in combination with interpolation-based or numerical error indicators.

Turbulent boundary layers have been studied extensively, both experimentally and computationally, and the mesh spacing requirements near the walls are well understood in terms of in-plane/lateral and thickness/normal resolution needs for different turbulence modeling approaches. These near-wall resolution requirements are usually defined in a dimensionless form of wall or plus units (e.g., $\Delta x^+$, $\Delta y^+$ and $\Delta z^+$) and vary according to the turbulence mode (e.g., RANS, LES or DNS) and the type of wall treatment (i.e., resolved or modeled). Since the mesh spacing requirements in the boundary layer region largely depend on the flow physics, it is advantageous to use this insight to drive the local adaptivity. Mesh resolution and structure in these regions need to be carefully chosen, with an emphasis on setting the wall normal spacing in a correct manner. Some strategies for selecting parameters of meshes inside the boundary layer region are explained in~\cite{Chitale}. 

In this paper we explore ways to set parameters for boundary layer meshes in the wall normal direction using the flow physics. Other regions of the mesh are adapted using a mesh size field derived from the Hessian of the flow solution quantities. We apply this technique to incompressible turbulent flows to study its effectiveness and efficiency.

\section{IMPLEMENTATION DETAILS}

\begin{figure}[h!]
\begin{center}
\begin{floatrow}
\ffigbox{
 \includegraphics[width = 8 cm]{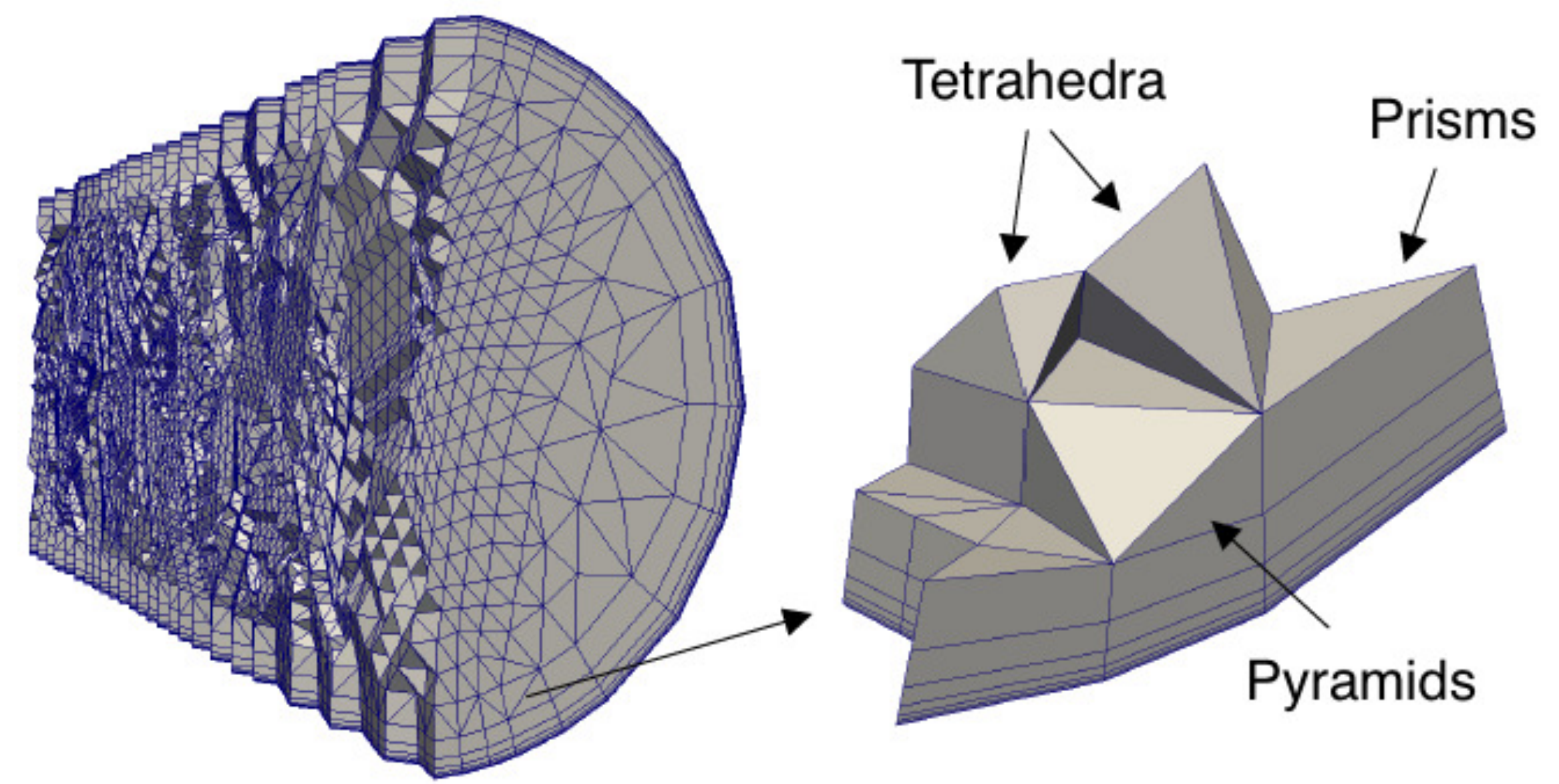}
}
{
\caption{Boundary layer mesh
 for a pipe geometry}
 \label{f:PipeBL}
 }
 \hspace{-0pt}
 \ffigbox{
%[] {
 \includegraphics[width = 8 cm]{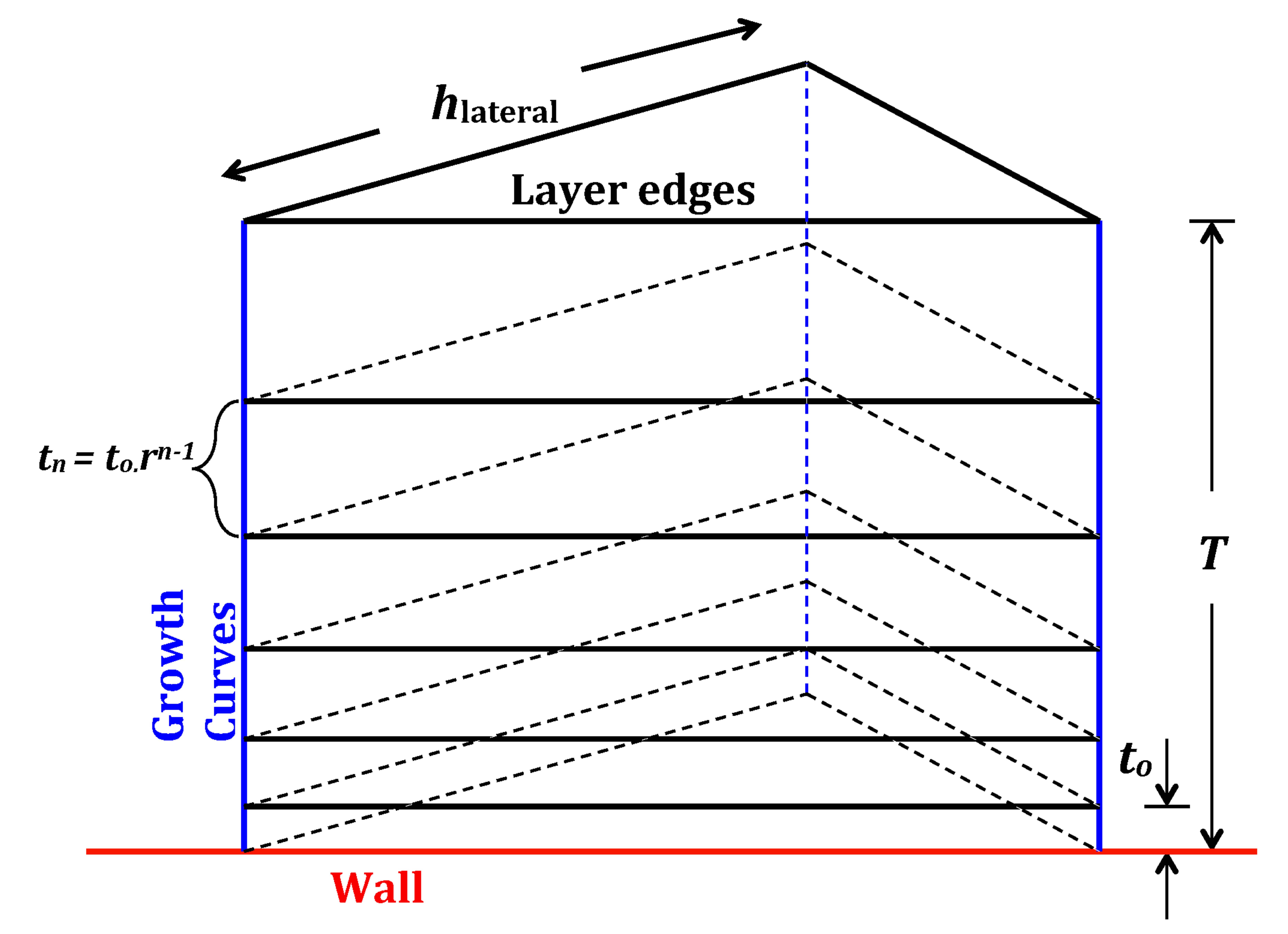}
}
{
\caption{Sketch of a boundary layer mesh}
 \label{f:BLStructure}
}
\vspace{-25pt}
 \end{floatrow}
 \end{center}
\end{figure}

Semi-structured boundary layer meshes are widely used in numerical simulation of wall-bounded turbulent flows. Such meshes provide an easy way to achieve anisotropic elements with aspect ratio of 10,000 or more, without creating poorly shaped elements which would severely influence and degrade the flow resolution. Figure~\ref{f:PipeBL} shows an example of a boundary layer mesh for a simple pipe geometry~\cite{Ovcharenko}. For 3-D meshes, the layered portion of the mesh is comprised of prismatic elements whereas the interior unstructured portion is meshed with tetrahedral elements. Furthermore, pyramid elements are also encountered at the interface between the structured and the unstructured region, where a quad face of a prism is exposed to the unstructured portion of the mesh. 

%\newpage
The boundary layer mesh contains a structure that can be decomposed into a product of a layer surface (2D) and a thickness (1D)~\cite{Sahni}.
Similarly, mesh attributes for mesh resolution needs include in-plane/lateral and thickness/normal components.
The in-plane resolution is prescribed in a similar fashion as the unstructured portion of the mesh. This controls the mesh composed of triangles located on the wall (and in each layer).
On the other hand, the important parameters related to the thickness of layered portion of the mesh are: 
\begin{enumerate}
\item First cell height or thickness (normal distance of the first mesh point off the wall surface): $ t_0$
\item Total height or thickness of the layers: $ T$
\item Total number of layers: $n_{layers}$
\item Growth factor in height of two layers: $r$
\end{enumerate} 

Typically layers are created with a geometric progression leading to interdependence of these parameters that is given by the following equation:
\begin{equation}
 \label{e:BLEq}
T = t_0 \sum_{i=1}^{i=n_{layers}} r^{(i-1)}
\end{equation}
The growth factor is the multiplication factor by which the height of a layered element increases with respect to that of the previous layer. Three of the above parameters can be chosen independently, and the fourth one is determined through Eq.~\ref{e:BLEq}. Figure~\ref{f:BLStructure} shows the general structure of a boundary layer mesh and shows some of the key mesh attributes. Out of these, $ t_0$ and $ T$ are physically the most important ones. As described in Section~\ref{sec:ThickAdapt}, we use flow physics information from the boundary layer to set these parameters.

\subsection {Hessian Driven Anisotropic Adaptivity}

Outside the boundary layer, the mesh can also be anisotropic. Since the level of anisotropy required outside the boundary layer is much less, general unstructured anisotropic meshes are used and the mesh anisotropy is defined using the well-known Hessian (or interpolation error) based methods~\cite{Buscaglia, Diaz, PerPei92, Pain01}. The anisotropic adaptivity used in this work, is based on local modifications of the mesh elements following a {\it mesh metric field}~\cite{XLi2}. The mesh metric is derived from a Hessian matrix, which is a symmetric matrix constructed from the second derivatives of particular flow solution variables. Traditionally, speed and density are chosen as the solution variables, but a combination of different variables can also be used.

The Hessian matrix is decomposed as $ H = R\Lambda R^T$, where $ R$ is the matrix of eigenvectors and $ \Lambda$ is the diagonal matrix of eigenvalues ($ \lambda$). The directions associated with the eigenvectors are referred to as the principal directions and the eigenvalues are equivalent to the second derivatives along these directions. High eigenvalues are associated with higher error in the corresponding direction. Similarly, a low eigenvalue means lower error in the corresponding direction~\cite{Sahni}. To achieve a suitable mesh resolution in different directions, a uniform distribution of local errors is applied in the principal directions which leads to $ h^2_k |\lambda_k| = \epsilon$, where $ \epsilon$ is a user specified tolerance for the error and $ h_k$ is the desired size in the $ k^\text{th}$ principal direction. More details of the size field computations can be found in Sahni et al.~\cite{Sahni2}.

\begin{figure}[h!]% order of placement preference: here, top, bottom
\begin{center}
\vspace{10pt}
\includegraphics[width=9 cm]{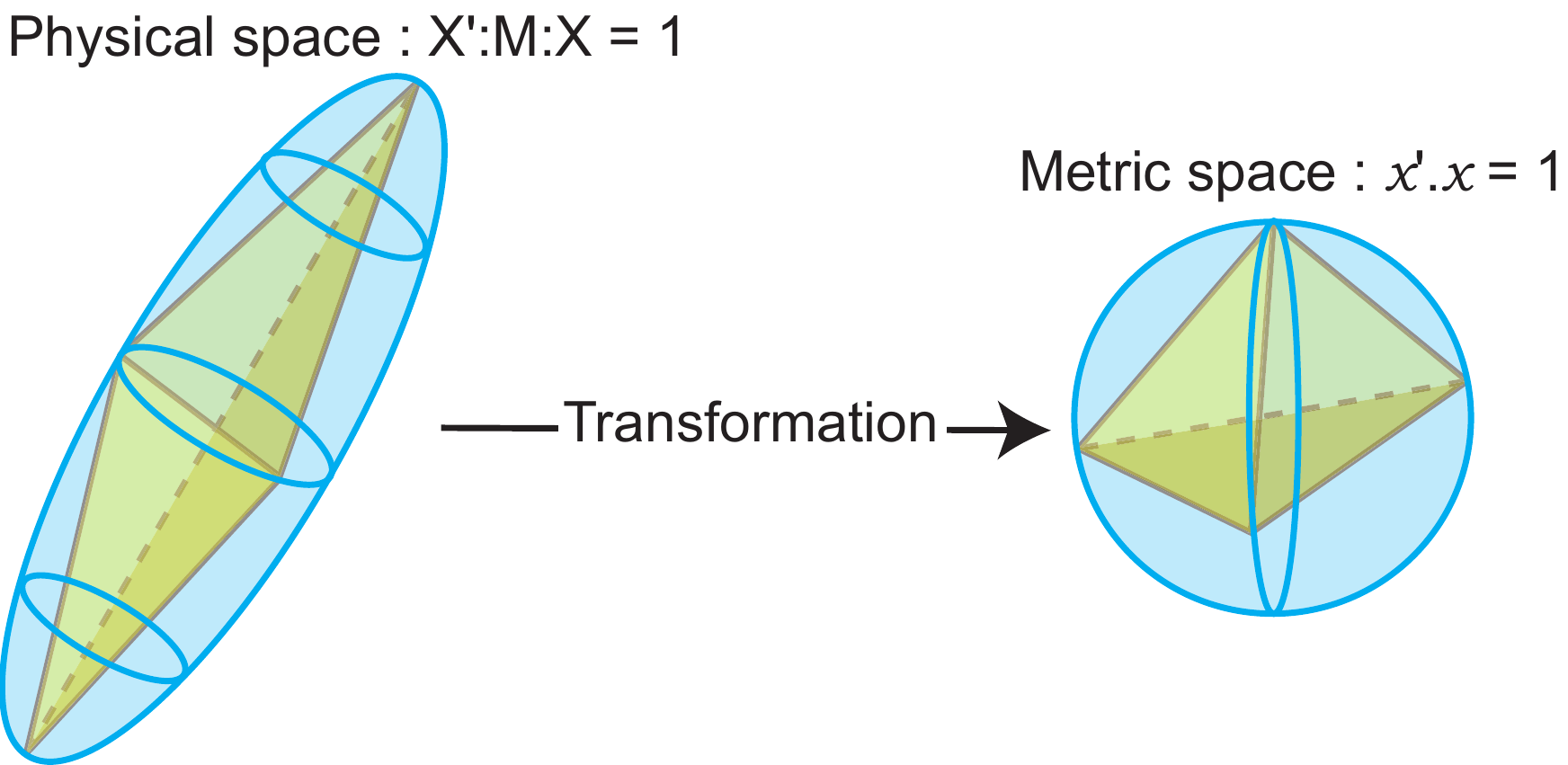}
\caption{Transformation associated with the mesh metric tensor~\cite{Sahni}} 
\label{f:MeshMetricTransform}
\end{center}
\vspace{-5pt}
\end{figure}

The mesh metric field can be thought of as a transformation matrix which defines a mapping of an ellipsoid in the physical space into a unit sphere in the metric space, as shown in Figure~\ref{f:MeshMetricTransform}. An element of any shape in the physical space is transformed to an equilateral element in the metric space with this transformation. The goal of the mesh adaptation software is to achieve unit edge lengths in the metric space. For meshes of complex domains, this criteria is usually relaxed to constrain edge lengths in the metric space to be within an interval close to 1~\cite{Diaz, XLi2}.

In this work, we use a commercial mesh generation and adaptation package provided by Simmetrix Inc.~\cite{Simmetrix}. The mesh generation process employs three basic steps. In the first step the surface mesh is created on a patch by patch basis using a general purpose anisotropic triangular mesh generator. Next, an advancing front procedure is used to produce the boundary layer meshes on the selected surface patches~\cite{Garimella}. The remainder of the domain is then filled by a general purpose anisotropic tetrahedral mesh generator. All steps in the meshing procedure interact with the original domain definition (e.g., CAD model) to ensure the correct geometric approximation of the mesh. Moreover, the volume meshing steps are allowed to introduce local modifications to the surface mesh if such modifications yield a better overall mesh. The mesh adaptation procedure employs a generalized set of mesh entity splits, collapses, swaps and compound operators to convert the given mesh to one that satisfies the anisotropic mesh metric field given. The two overall steps in the mesh adaptation procedure are (i) adapting the boundary layer such that the boundary layer structure is maintained~\cite{Sahni} and (ii) adapting the remaining interior mesh~\cite{XLi2}. Both of these steps interact with the original domain definition to ensure the correct geometric approximation of the mesh.
   
\subsection{Extension to Boundary Layers}

The methodology outlined above works well for unstructured elements. When working with boundary layers we want to preserve their structured nature, and using this technique directly does not guarantee that. To extend anisotropic adaptivity to boundary layer meshes we instead use the approach described below.

\begin{figure}[h!]% order of placement preference: here, top, bottom
\begin{center}
\vspace{10pt}
\subfigure[Decomposition of the boundary layer mesh] {
\includegraphics[width=7.2cm]{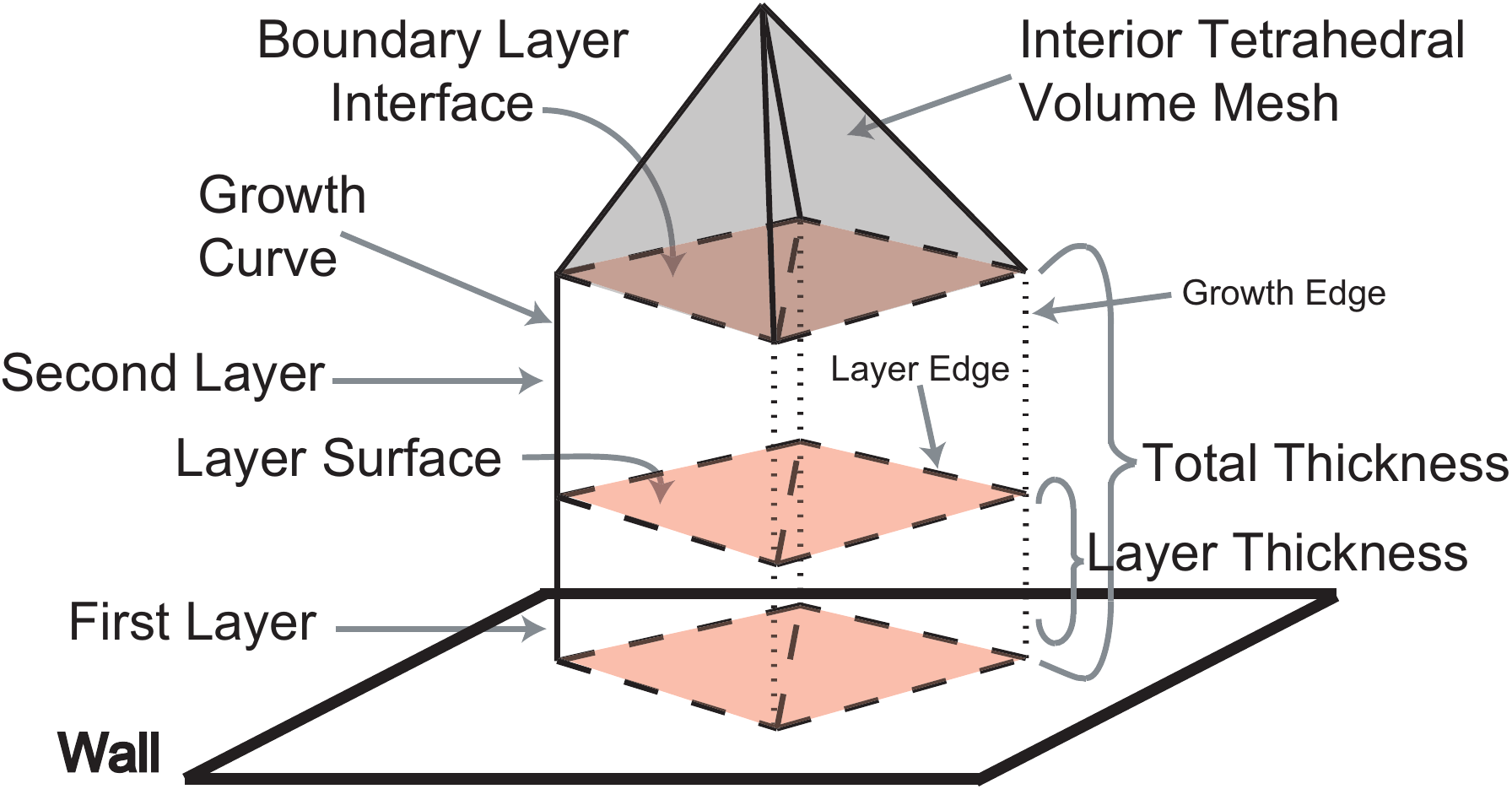}
\label{f:BLDecompose}
}
\subfigure[Decomposition of the ellipsoid]{
\includegraphics[width=7.4cm]{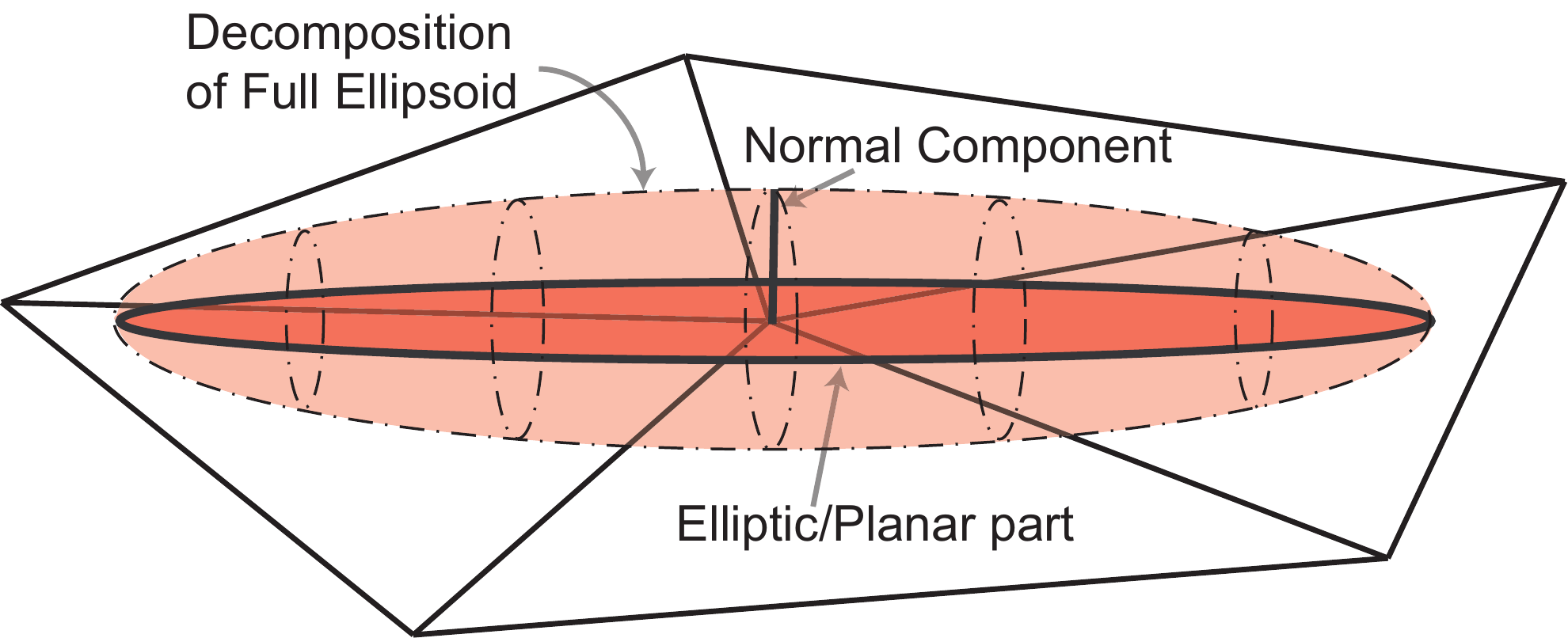}
\label{f:EllipsoidDecompose}
}
\end{center}
\vspace{-5pt}
 \caption{Conceptual extension of the Hessian approach to boundary layers~\cite{Sahni}}
 \label{f:MeshMetricAndEquation}
\end{figure}

Figure~\ref{f:BLDecompose} shows a conceptual decomposition of the boundary layer mesh. The boundary layer meshes can be viewed as a product of a layer surface (2D) and a thickness (1D) mesh. The lines which are orthogonal to the wall are referred to as the {\it growth curves}, and the triangular surfaces parallel to the wall are referred to as the {\it layer surfaces}. Each layer of elements is formed with the help of the layer surfaces above and below, connected by the growth edges in between. The mesh size on the layer surfaces is referred to as the {\it in-plane} or lateral size and that on the growth curves is referred to as the {\it normal} spacing or thickness. The ellipsoid in Figure~\ref{f:MeshMetricTransform} can be decomposed as an ellipse projected on the layer surface and a normal component aligned with the growth curve. This concept is shown in Figure~\ref{f:EllipsoidDecompose}. 

Adaptivity is carried out in two stages; {\it in-plane adaptation} that achieves the required mesh sizes on the layer surfaces and does not affect the thickness, and {\it thickness adaptation} which changes the normal spacing of the boundary layers. The in-plane adaptation is driven by the mesh metric field calculated from the Hessian as described in this section (see~\cite{Sahni} for more details). The thickness adaptation is driven by the procedures outlined in the next section. % which are also explained in Chitale et al.\cite{Chitale}. 

%%%%%%%%%%%%%%%%%%%%%%%%%%%%%%%

\subsection{Adaptive Control of Layer Thickness}
\label{sec:ThickAdapt}

To efficiently resolve the boundary layers in turbulent flows, careful control is needed for the distribution of points in the wall-normal direction. The Hessians tend to be less accurate near the walls and therefore, they are not a good candidate to drive thickness adaptation for such a critical flow region. Since, the mesh spacings in this region are largely dictated by the boundary layer profile and the turbulence model being used, this information must be used for thickness adaptation.

%\begin{wrapfigure}{R}[-9mm]{0.5\textwidth}
\begin{figure}[h!]% order of placement preference: here, top, bottom
%\vspace{-20pt}
\begin{center}
	 \includegraphics[width = 14.0 cm]{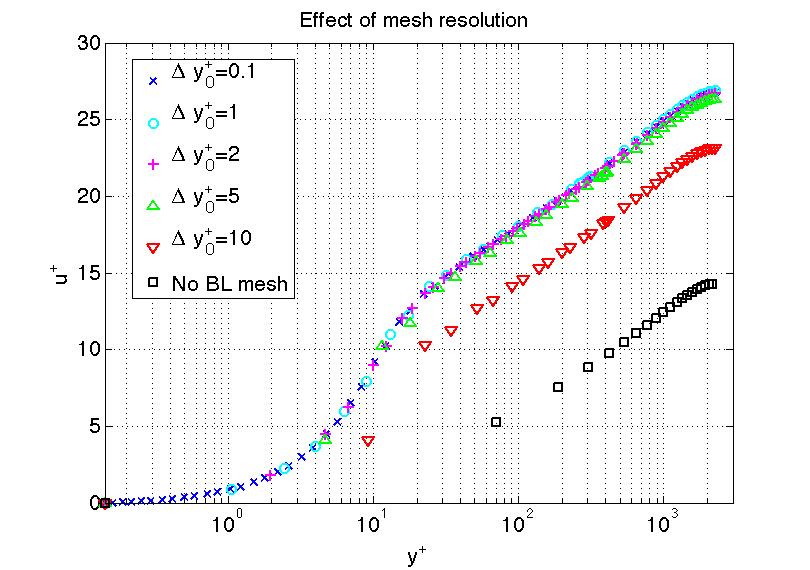}
%\vspace{-10pt}
 \caption{Effect of the first cell height on a turbulent boundary layer profile as predicted from the Spalart-Allmaras (RANS) turbulence model}
 \label{f:PipeBLProfile}
 \end{center}
 %\vspace{-10pt}
 \end{figure}

%\end{wrapfigure}	

To demonstrate how critical the distribution of points normal to the wall is, Figure~\ref{f:PipeBLProfile} shows boundary layer profiles for a turbulent pipe flow. The first cell height is varied from $ t^+_0$ or $\Delta y^+_0$ of $0.1$ to $10$. The total height ($ T$) and the number of layers ( $n_{layers}$) are kept constant. The results show inaccurate boundary layer profiles when $\Delta y^+_0 > 5$, with the worst profile obtained with no boundary layer mesh. These results were obtained with the Spalart-Allmaras ~\cite{SA} RANS turbulence model (RANS-SA model) without any wall modeling, and with linear stabilized finite elements. This behavior would be different for other choices (e.g., turbulence model or wall modeling), however, when $\Delta y^+_0$ is above a certain value similar trend is expected.

\subsubsection{Types of boundary layers in flows}

\begin{figure}[h!]
\begin{center}
\subfigure[Laminar boundary layer] {
 \includegraphics[width =4.5 cm]{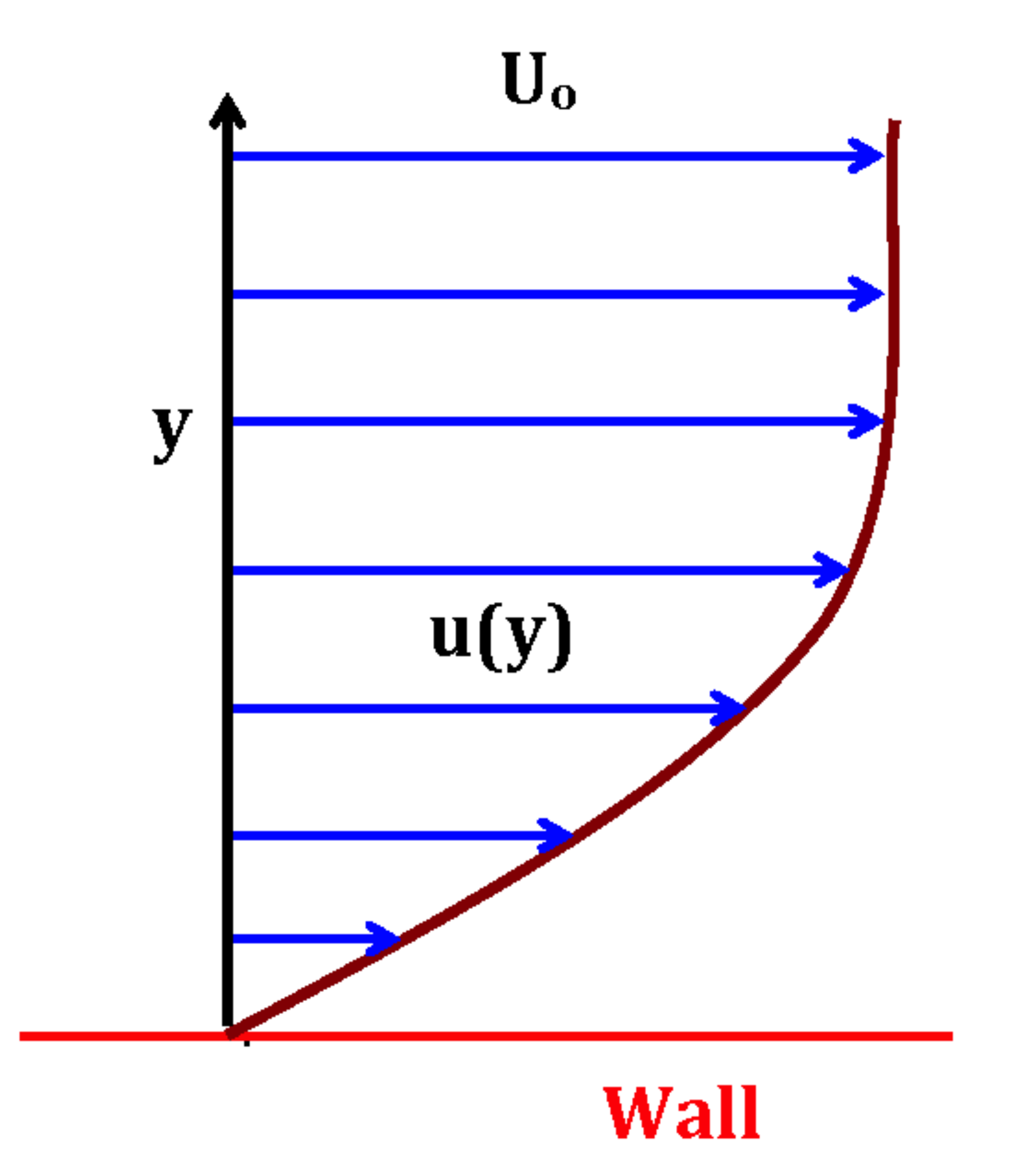}
 \label{f:BL_Laminar}
}
\subfigure[Turbulent boundary layer] {
 \includegraphics[width = 4.5 cm]{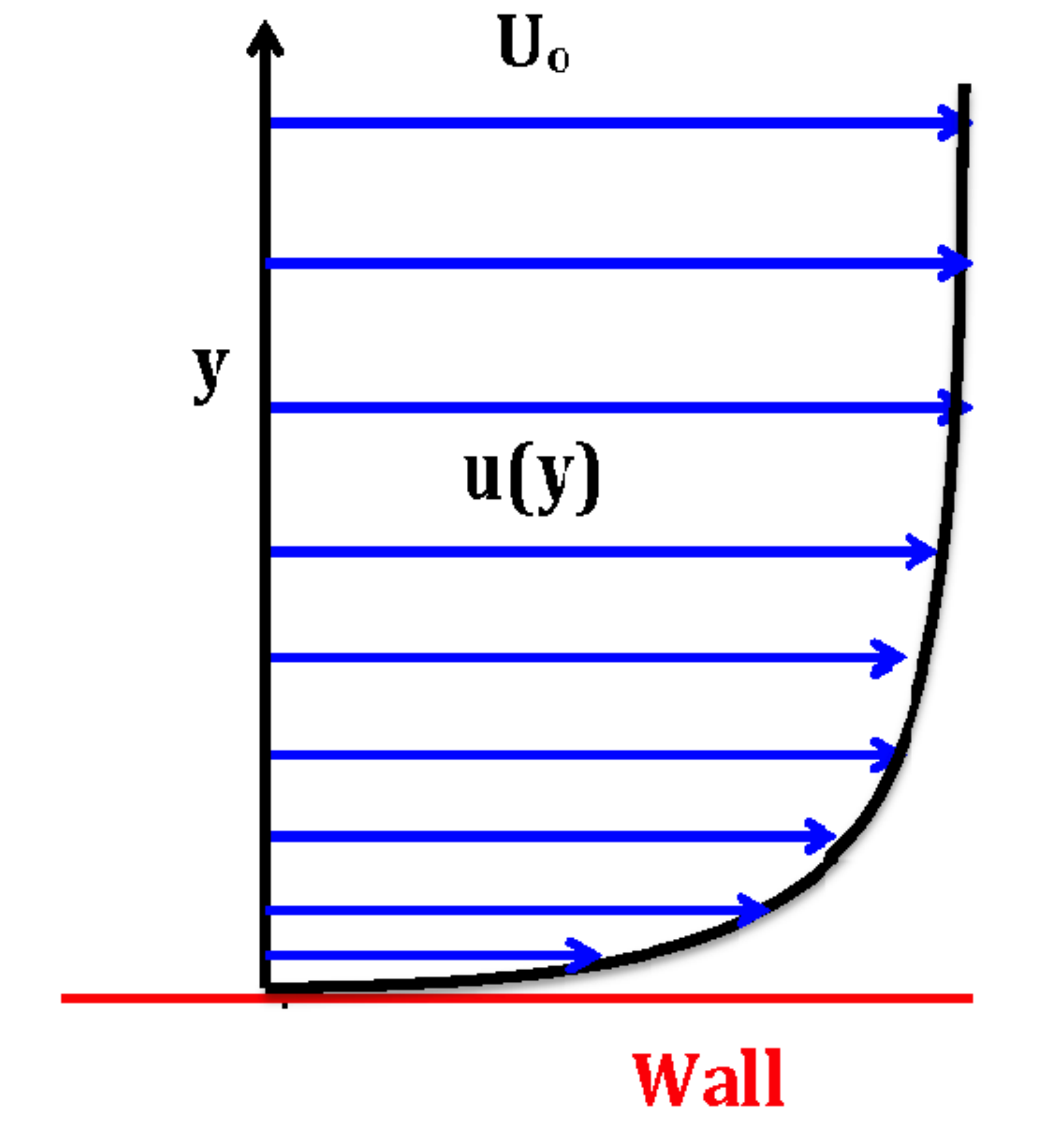}
 \label{f:BL_Turbulent}
 }
 \subfigure[Separated boundary layer] {
 \includegraphics[width = 4.5 cm]{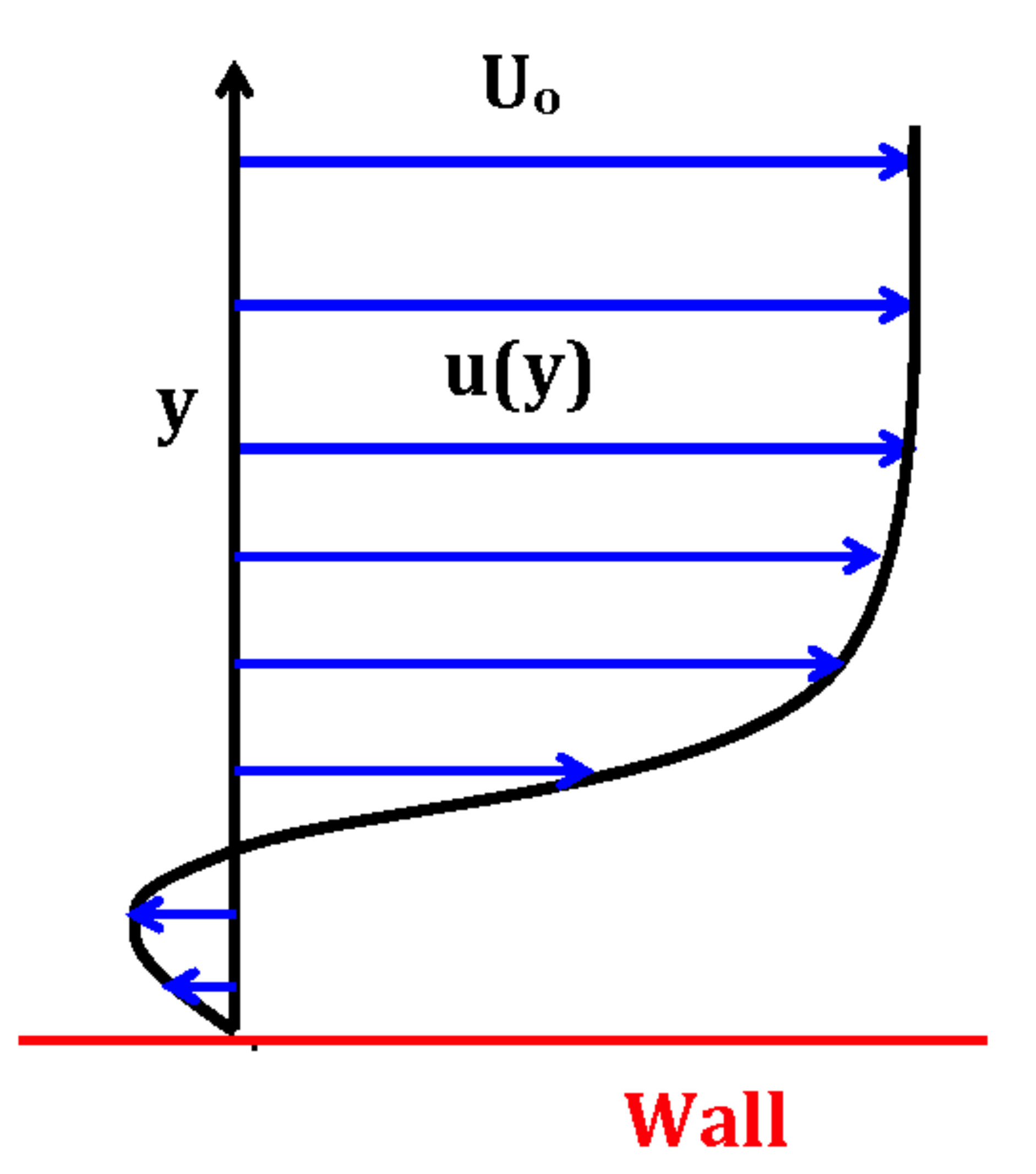}
 \label{f:BL_Separated}
 }
 \caption{Types of boundary layer profiles}
 \label{f:BLProfiles}
  \end{center}
  \vspace{-15pt}
\end{figure}

At this point it is important to define the different classes of boundary layers in the flow. This is because control of the mesh parameters is different for each type. The usual classification includes laminar and turbulent boundary layers. However, this paper focuses on turbulent boundary layers as they are the most prevalent type for high Reynolds number flows and are much more complex to deal with in numerical simulations. Also, the mesh spacing requirements to resolve the turbulent boundary layers are much tighter than that for the laminar boundary layers; in the former much steeper profiles or larger velocity variations are experienced near the wall (see Figure~\ref{f:BLProfiles}).
% *** not needed to say this (this may be sufficient for laminar cases but if we use the same strategy then we will not be optimal for laminar cases)*** Thus our treatment for turbulent boundary layers is sufficient for resolving laminar boundary layers as well. 

The second classification of the boundary layers relates to whether the boundary layer is attached or separated. In many flows, due to adverse pressure gradients and/or sharp turns and corners, boundary layers separate from the wall and form a separated or free shear layer. The treatment of separated boundary layers needs special care as the flow physics and mesh resolution needs in this region are different than that of the attached boundary layers. Figure~\ref{f:BLProfiles} shows the two types of attached boundary layer profiles and a typical separated boundary layer profile. In both classifications, we use local wall shear stress to incorporate flow physics.

\subsubsection{Calculation of the wall shear stress}
\label{s:shear}

The near-wall mesh spacing requirements for turbulent models depend on distance from the wall in wall units ($y^+$), which needs the knowledge of the wall-friction velocity $ u_\tau$. The friction velocity can be typically calculated from the wall shear stress $ \tau_w$ as $ u_\tau = \sqrt{\frac{\tau_w}{\rho_w}}$. Many solvers readily provide wall shear stress as a field after a post-processing step based on local gradient at the walls (e.g., finite difference or element-based gradient values). However, this typically results in numerical noise due to involvement of velocity derivative. We use an alternative method to calculate this field in a fast and an accurate manner.

Since boundary layers in high Reynolds number flows are mostly turbulent, one method to calculate $ \tau_w$ is by using the Spalding's law. It gives $ y^+$ (dimensionless wall distance) as a function of $ u^+$ (dimensionless velocity), written in an implicit form given by~\cite{Spalding}:
\begin{equation}
\label{e:SpaldingLaw}
y^+ = f(u^+) = u^+ + A[e^{(\kappa u^+)}-1-(\kappa u^+)^2/2 - (\kappa u^+)^3/6 - (\kappa u^+)^4/24]
\end{equation}

where $A=0.1108$ and $\kappa = 0.4$ are dimensionless constants. $ u^+$ is the dimensionless (mean) velocity along the boundary layer and is obtained by normalizing the local velocity ($u$) by the friction velocity ($u_\tau$), i.e., $u/u_\tau$. This law is valid through the inner layer region of an attached turbulent boundary layer on smooth surfaces and is therefore useful for wall shear stress computation since it requires information from a region very near to the wall (i.e., sublayer or buffer region of the turbulent boundary layer). The case for separated flow is further described in Section~\ref{s:SepBL}. Using Eq.~\ref{e:SpaldingLaw}, $ u^+$ and thus $ u_\tau$ can be calculated at various points along the growth curve, with following iterative approach (discussed for a given point).

\begin{enumerate}
\item Calculate the distance of the given point on the growth curve from the wall: $\Delta y $.
\item Retrieve the velocity at this point from the flow solution: $u$.
\item Guess an initial value of $ u^1_\tau$ and calculate initial $ u^+ = u/u^1_\tau $ and $\Delta y^+ = \Delta y u^1_\tau / \nu_w$. 
\item Use Newton's method to iteratively solve Eq.~\ref{e:SpaldingLaw} until convergence is reached (to a specified tolerance) and update value of $ u_\tau$ (at each iteration).
\end{enumerate}

This process can be repeated at any number of points on the growth curve. In the end, an average over the points gives the final friction velocity $ u_\tau$ at the wall point of the growth. We typically use first 3 to 5 points along the growth curves to calculate $ u_\tau$. The wall shear stress can then be calculated as $ \tau_w = u_\tau^2 \rho_w$. This entire process is done for each growth curve to get $ \tau_w$ at each wall vertex.

Another method to calculate a quick and an approximate estimate of the wall shear stress, is using a finite difference approach near the wall. Using the first vertex from the wall and known $ u$ and $ \Delta y_0$ values at that point, $\tau_w $ can be calculated using $ \tau_w \simeq \mu \frac{du}{dy} \simeq \mu \frac{u}{\Delta y_0}$.  At the first point off wall, $ du$ equals $ u$ because the velocity is zero at the walls in the current cases. This alternative method is used for flow regions where the boundary layers are not attached and Spalding's law is not applicable.

\subsubsection{First cell or layer height ($ t_0$)}

As mentioned before, different turbulence modeling approaches have different mesh spacing requirements close to the walls. Even in the same family of turbulence models such as RANS, different approaches require varying mesh spacings depending on which specific turbulent model is used and whether the boundary layer is integrated to the wall (wall resolved approach) or if wall functions are used (wall modeling approach). The wall resolved approach makes a low Reynolds number assumption near the walls and requires that the first cell height is inside the viscous sublayer of the boundary layer ($ \Delta y^+_0 \leq 5$). The wall modeling approach makes suitable assumptions for near wall behavior of the boundary layer and requires that the first cell height is beyond the viscous sublayer and into the log-layer or overlap region ($ \Delta y^+_0 > 30$). If these requirements are not met for either of these modeling classes, then large numerical errors are incurred in turbulence calculations predicting erroneous behavior, as seen in Figure~\ref{f:PipeBLProfile}. However, the friction velocity, which is required to calculate $ \Delta y^+_0$, is not known {\it a priori}. This makes adaptive control of the first cell height very important. 

Let us assume that the turbulence model requires first cell height to be equal to $ (\Delta y^+_0)_{req}$. If we have an initial coarse mesh with a computed solution, using the wall shear stress ($ \tau_w$), desired value of $ t_0$ (in a local fashion) can be calculated, by the following algorithm:

\begin{enumerate}
\item Get the local kinematic viscosity ($\nu_w $) and desired $ (\Delta y^+_0)_{req}$ according to the turbulence model from the user (suggested values are $ (\Delta y^+_0)_{req}= 1-5 $ for wall resolved RANS-SA, $ (\Delta y^+_0)_{req} = 30-50$ for wall modeled $k-\epsilon$, 0.5 for wall resolved $ k-\epsilon$ etc.) 
\item Calculate the friction velocity $ u_\tau$ as discussed in Section~\ref{s:shear}
\item Calculate the first layer height of the boundary layer by $ t_0 = {\nu_w (\Delta y^+_0)_{req}}/u_\tau$. 
\end{enumerate}

\subsubsection {Total height of the layered mesh ($T$)}
\label{s:BLHeight}

It is desirable to have the total height of the layered portion of the mesh to be equal to or greater than the height of velocity boundary layer given by $ \delta_{99}$. $ \delta_{99}$ is the distance from the wall at which the velocity becomes 99\% of the local free stream velocity (i.e., at the edge of the boundary layer), and is an accepted measure of the boundary layer thickness. It is usually difficult to calculate $\delta_{99}$ directly as it requires knowledge of a reference velocity. For simple problems (e.g., a flat plate), the reference velocity is usually the constant free stream velocity, but it can have significant local variations for problems of interest, where the flow as a whole undergoes local acceleration or deceleration. This presents a difficulty in directly calculating the boundary layer height.

To calculate $ T$, we base our approach on the observation that vorticity outside of an attached boundary layer is negligible. Since the boundary layers have the largest velocity gradients very close to the wall, vorticity in this region is the highest and decreases as one moves farther away from the wall. As boundary layer growth curves are perpendicular to the wall (or close to perpendicular), one can walk along these edges starting from the wall, and determine the point at which the vorticity drops below a threshold value. This threshold value depends on the local maximum value of vorticity for attached boundary layers, which is most often encountered at the wall. In our analysis, we have found that a good value for the threshold is 0.02\% of the wall vorticity magnitude.

\subsubsection{Growth factor ($r$) and number of layers ($n_{layers}$)}

To increase the height of the boundary layer elements away from the wall, a growth factor (also known as the stretching factor) greater than 1 is used. This is because the tightest mesh spacing is required very close to the wall, but this requirement is not as strict further away from the wall. An ideal scenario would be to achieve the height of the last layer equal to the mesh sizes in the unstructured region of the mesh in order to get a smooth transition. The mesh adaptation process provides the option of a boundary layer gradation factor, which controls the smoothness of transition of boundary layer into the unstructured part of the mesh. 

There are general guidelines for what the desired growth factor should be, from the perspective of turbulence modeling. Spalart~\cite{SpalartTrends, SpalartDES} states that the growth factor should be close to $1.25$ to accurately capture the log-layer. Generally a growth factor beyond the value of $1.4$ is deemed too large for accurately capturing the boundary layers. Many meshing tools are based on setting $ t_0$, $ T$ and $ n_{layers}$, and the growth factor is automatically calculated, internally. The accuracy then in turn hinges on the knowledge and prior calculation to make sure that the growth factor being calculated is acceptable. 

The adaptation tool gives the ability to set the growth factor at each wall vertex. We set $ r$ in the range of $1.2-1.25$ to be within the acceptable limits. Selecting a growth factor less than $1.2$ has the disadvantage of creating more elements than needed. The number of layers are then calculated using Eq.\ref{e:BLEq}.

\subsubsection{Developments for separated boundary layers}
\label{s:SepBL}

The techniques described above for calculating the different aspects of the boundary layer meshes work well for attached boundary layers. However, separated boundary layers need extra care and special detection strategies due to different flow physics that must be captured. 

To treat separated boundary layers properly, they must first be detected. As it can be seen from Figure~\ref{f:BLProfiles}, they have a unique profile characterized by flow reversal. We again make use of the wall normal growth curves and walk along the growth edges to detect a change in the flow direction. If a change (usually more than $120^\text{o}$) is detected across the profile, then the vertex on the wall is marked as {\it separated}; otherwise the boundary layer is treated as an attached boundary layer. This method requires that the total height of the layers in this region at least exceeds the height at which the flow direction is reversed. This means that typically initial boundary layer meshes for such regions should be tall enough and mesh very close to the wall should be fine enough to capture the flow reversal. Currently we make sure that this criteria is satisfied through initial meshing, but an iterative adaptive procedure like the one we use eventually leads to suitable meshes which are able to capture this effect. 

The method of using Spalding's law to calculate the wall shear stress, is not appropriate for separated boundary layers, since the typical turbulent profile is absent. In such a case, where separation is detected, wall shear stress is calculated using the finite difference method explained earlier. The accuracy of such calculations is not as good as other methods, but it gives a reasonable estimate. Also, for separated boundary layers, the first cell height of the layer is not as crucial as for the attached boundary layers, hence such an approximate approach is justifiable. 

For separated boundary layers, the free shear layer might get separated from the wall to a fair distance, in which case it might not be prudent to increase the boundary layer height. Even though it would be a good feature to separate or detach the layered mesh from the wall and instead follow the free shear layers in order to resolve them effectively, this capacity is still under development. The technique explained in Section~\ref{s:BLHeight} for attached boundary layers predicts that the boundary layer's height should be increased to the height of the complete shear layer. This means that the boundary layer should grow from the wall into the separated shear layer. However, this is not always practical for separated boundary layers as this height might introduce excessive stretching of the elements near the interface. In a more practical approach, the boundary layer height is maintained beyond the height at which the flow reversal is detected so that the boundary layer mesh is tall enough to ``sense" the change in the flow direction. Beyond this height, rest of the region is meshed with unstructured elements, with specific care to resolve the free shear layer. This approach is used in this work. For anisotropic adaptivity, velocity Hessians give good resolution in these layers since the anisotropy of the top of the shear layer is not very high. 

\section{RESULTS}

This section summarizes results for 2 incompressible flow cases that we have tested our approach on. The first case is a simple turbulent flat plate and the second case is a NACA 0012 airfoil.

\subsection{Turbulent Flat Plate}

The simplest example to test the changing boundary layer thickness for incompressible flows is a turbulent flat plate.  The case setup has a Mach number of 0.2 and a Reynolds number of 10 million based on the total length of the plate, which is 10$ m$. The case is run as a steady case with wall resolved RANS-SA turbulence model. This turbulence model needs the $ \Delta y^+_0$ to be between 1 to 5. We set it to 1 for our calculations, to have an added factor for safety. 
 
\begin{figure}[h!]
\begin{center}
    \includegraphics[width = 12cm]{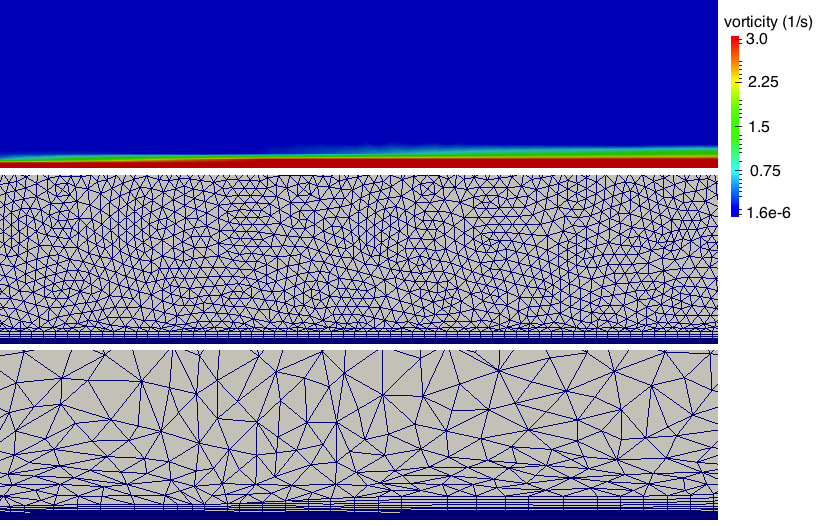}
\end{center}
 \caption{Change in the boundary layer thickness of the flat plate}
 \label{f:FlatPlateVort}
\end{figure}

Figure~\ref{f:FlatPlateVort} shows vorticity magnitude, the initial mesh and the adapted mesh for the flat plate. The increasing boundary layer height is clearly seen by looking at the vorticity magnitude. The initial coarse mesh has a constant thickness boundary layer mesh. The adapted mesh in the boundary layer, clearly shows increasing total boundary layer thickness of the mesh which roughly follows the behavior of vorticity magnitude. 

\begin{figure}[h!]
\begin{center}
    \includegraphics[width = 10cm]{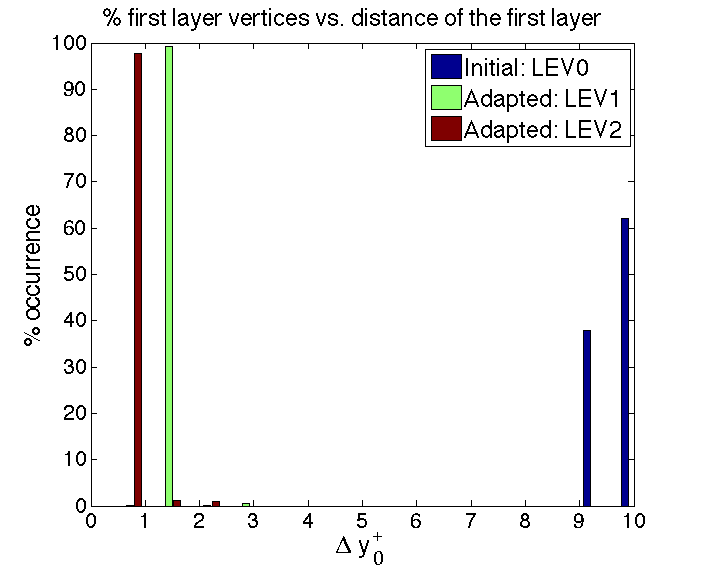}
\end{center}
 \caption{\% occurrence of vertices vs. $ \Delta y^+_0$}
 \label{f:FlatPlate_yplus}
\end{figure}

The first cell height of the boundary layer mesh also changes with adaptivity but cannot be seen in the mesh pictures due to the small value it takes. Figure~\ref{f:FlatPlate_yplus} quantitatively presents the change in the first cell height. The figure plots \% of vertices against $ y^+$ values, where vertices are the first points off the wall in the boundary layer mesh. For the initial coarse mesh, most of the first cell vertices lie at $ y^+$ value greater than 9. This is not optimal for the turbulence model that we use as the recommended value is $ y^+ = 1$. As the mesh is adapted, the calculations explained in Section~\ref{sec:ThickAdapt} predict the right $ t_0$ for out required $ \Delta y^+_0$ of 1. The adjustment of $ t_0$ can be viewed from this plot as most of the first cell vertices (nearly 98\%) now lie between $y^+$ of 1 and 2 for the first adapted mesh and very close to 1 for the second adapted mesh. 

\begin{figure}[h!]
\begin{center}
\subfigure[$ C_f$ for the flat plate] {
    \includegraphics[width = 8.1cm]{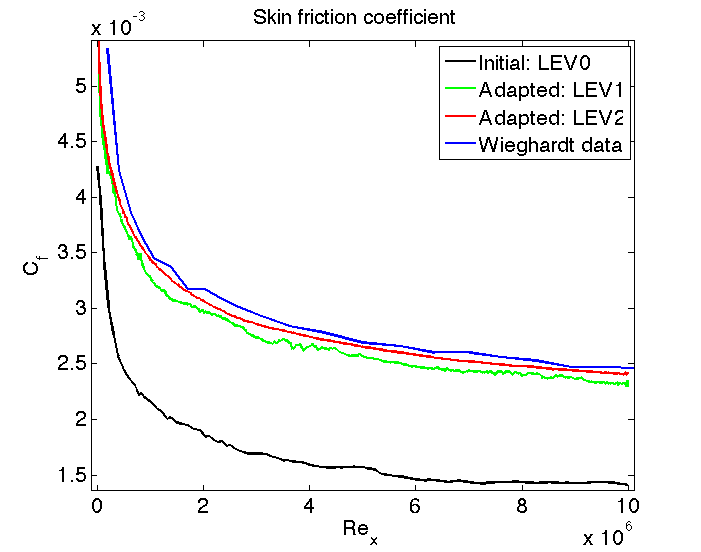}
     \label{f:FlatPlate_Cf}
 }
 \hspace{-10mm}
 \subfigure[BL height for the flat plate] {
    \includegraphics[width = 8.1cm]{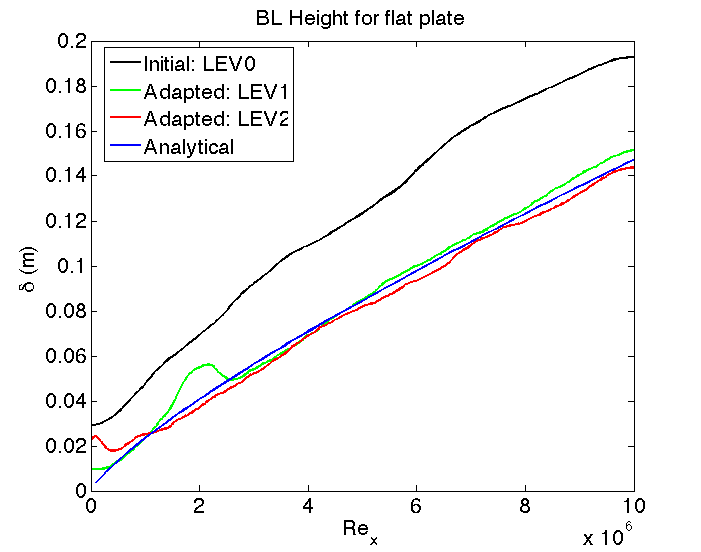}
     \label{f:FlatPlate_BLHeight}
 }
\end{center}
\vspace{-5mm}
 \caption{Calculations for the boundary layer}
 \label{f:FlatPlate_misc}
\end{figure}

However, these statistics also depend on the accuracy of calculating the wall shear stress. To check that it is under permissible limits, Figure~\ref{f:FlatPlate_Cf} plots skin friction coefficient along the plate length. $ C_f$ is calculated from the wall shear stress ($\tau_w$) as $ \tau_w/0.5 \rho U^2$. The initial coarse boundary layer mesh predicts $ C_f$ which shows similar behavior as the Weighardt experimental data~\cite{Weighardt}, but predicts significantly lower values than the experiments. As the boundary layer thickness is adjusted, the adapted meshes give a much closer $ C_f$ values to experimental data.   

Figure~\ref{f:FlatPlate_BLHeight} shows the change in boundary layer height calculation along the length of the flat plate. The calculations are compared with analytical boundary layer height calculated by $ 0.377x/(Re_x)^{(1/5)}$, which is derived from the 1/7th power law profile. The initial mesh over predicts the boundary layer heights almost for the entire length of the mesh. As the thickness of boundary layer is adjusted, the adapted meshes show a much closer agreement with the analytical values. This shows that as the mesh and the boundary layer thickness are adapted we get more accurate results for both wall shear stress and boundary layer height. 

\begin{figure}[h!]
\begin{center}
\subfigure[$ Re_x =$ 1.0 million] {
    \includegraphics[width = 3.8cm]{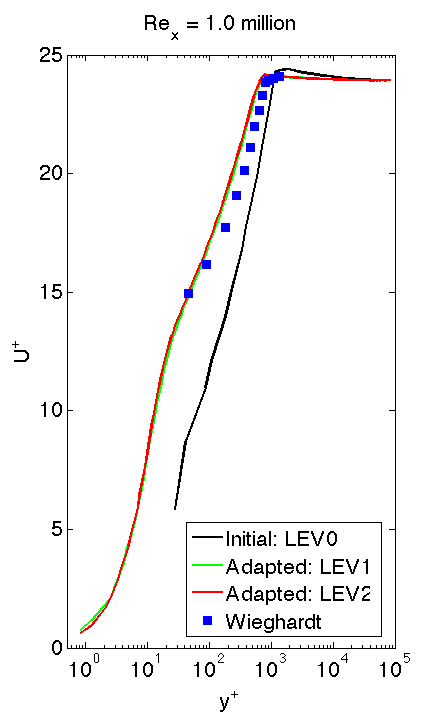}
     \label{f:FlatPlate_BL1Mil}
 }
 \hspace{-5mm}
 \subfigure[$ Re_x $= 2.7 million] {
    \includegraphics[width = 3.8cm]{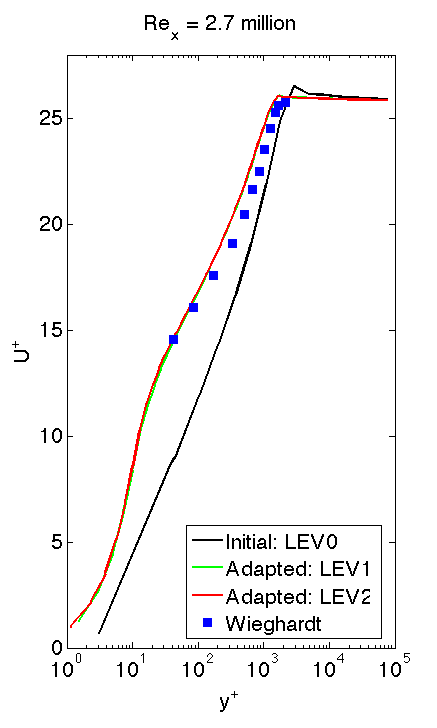}
     \label{f:FlatPlate_BL2Mil}
 }
  \hspace{-5mm}
 \subfigure[$Re_x$ = 5.0 million] {
    \includegraphics[width = 3.8cm]{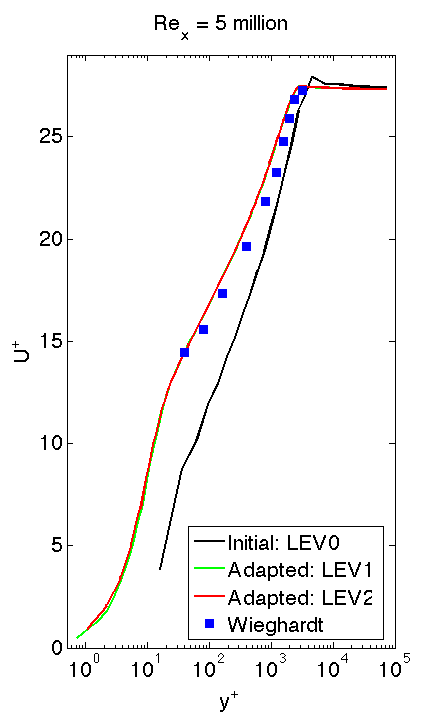}
     \label{f:FlatPlate_BL5Mil}
 }
  \hspace{-5mm}
 \subfigure[$ Re_x =$ 7.6 million] {
    \includegraphics[width = 3.8cm]{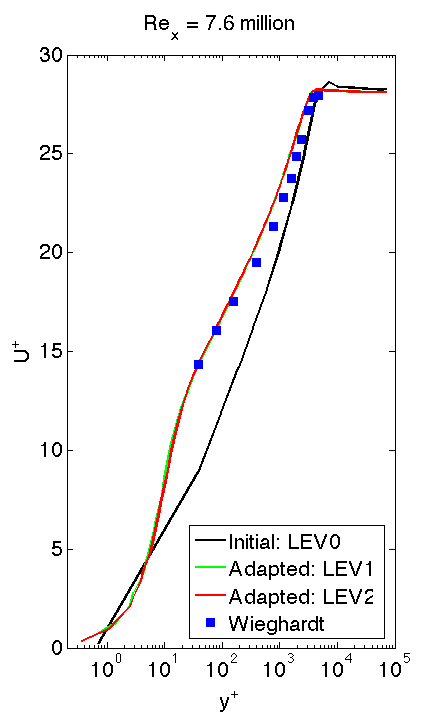}
     \label{f:FlatPlate_BL7Mil}
 }
\end{center}
\vspace{-5mm}
 \caption{Boundary layer profiles at different locations along the plate}
 \label{f:FlatPlate_BLProfiles}
\end{figure}

To determine the accuracy with which the boundary layer profiles are captured, we plot 4 boundary layer profiles at different locations on the flat plate in Figure~\ref{f:FlatPlate_BLProfiles}. The profiles are plotted in non-dimensional quantities, $ U^+$ and $ y^+$ and the x-axis is on a logarithmic scale. The experimental data is taken from Weighardt et al.~\cite{Weighardt}. The first cell height for the initial mesh is at $ y^+$ greater than 10, which is not ideal for the turbulence model we are using. This results in poor capturing of the boundary layer profile, even in the log layer. This shows that having some points in the viscous sub-layer is crucial for capturing of the overall profile for the Spalart-Allmaras turbulence model. Since the initial mesh does not satisfy this constraint, the velocity profiles it captures do not agree with the experiments. As the boundary layer mesh is adapted, we get enough points which are close to $ y^+$ of 1, giving sufficient resolution in the viscous sub-layer. This improves the velocity predictions in the log layer that agree well with the experimental data. The green and red curves are nearly identical, indicating that as the mesh is adapted further, the solution does not change significantly. 

\subsection{NACA 0012 Airfoil}

The second application is a NACA 0012 airfoil, which is a 2D airfoil geometry. Reynolds number based on the chord is 6 million. The Mach number is 0.15 and the flow is modeled with an incompressible flow solver. Three different angles of attack were studied and the initial mesh was adapted in each case with two adaptation cycles. The thickness of the boundary layers was calculated with the methodology given in Section~\ref{sec:ThickAdapt}. Hessians of time averaged scaled total pressure were used as error indicators to adapt the mesh in the unstructured region. The same initial mesh was used in all cases.

\begin{figure}[h!]
\begin{center}
\subfigure[LEV0 Initial mesh] {
    \includegraphics[width = 7.5cm]{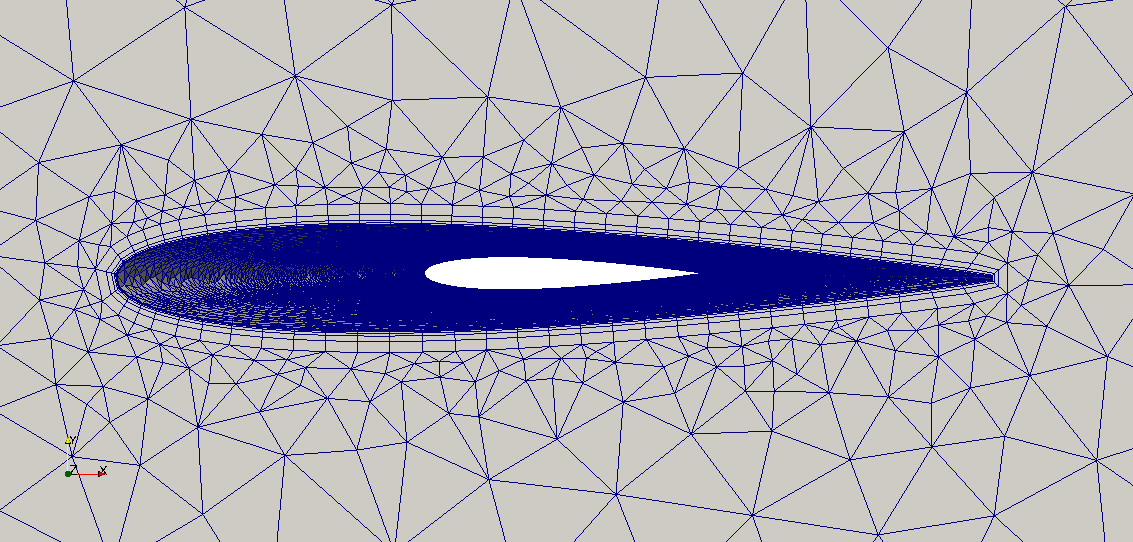}
     \label{f:NACA_InitMesh}
 }
 \hspace{-5mm}
 \subfigure[LEV2 adapted mesh: 0\degrees \hspace{0.1mm}  AoA] {
    \includegraphics[width = 7.5cm]{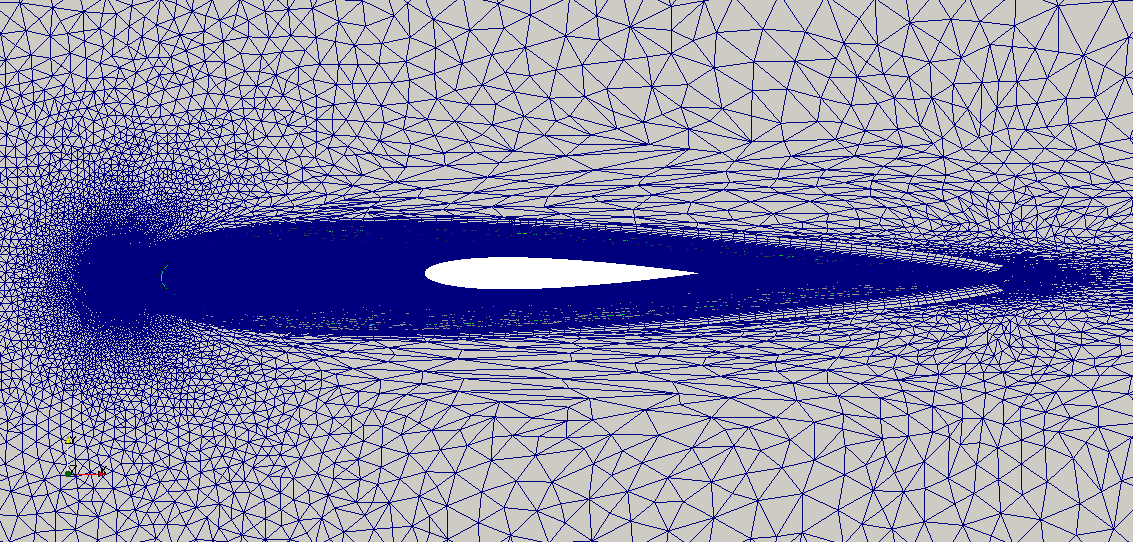}
     \label{f:NACA_0DegMesh}
 }
 
 % \hspace{-5mm}
 \subfigure[LEV2 adapted mesh: 5\degrees \hspace{0.1mm}  AoA] {
    \includegraphics[width = 7.5cm]{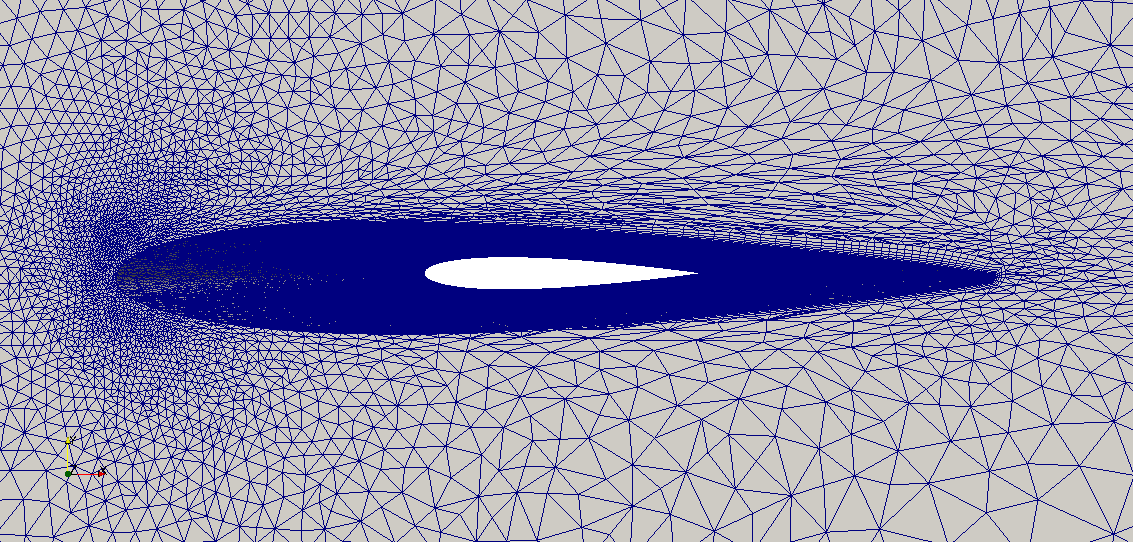}
     \label{f:NACA_5DegMesh}
 }
  \hspace{-5mm}
 \subfigure[LEV2 adapted mesh: 10\degrees \hspace{0.1mm}  AoA] {
    \includegraphics[width = 7.5cm]{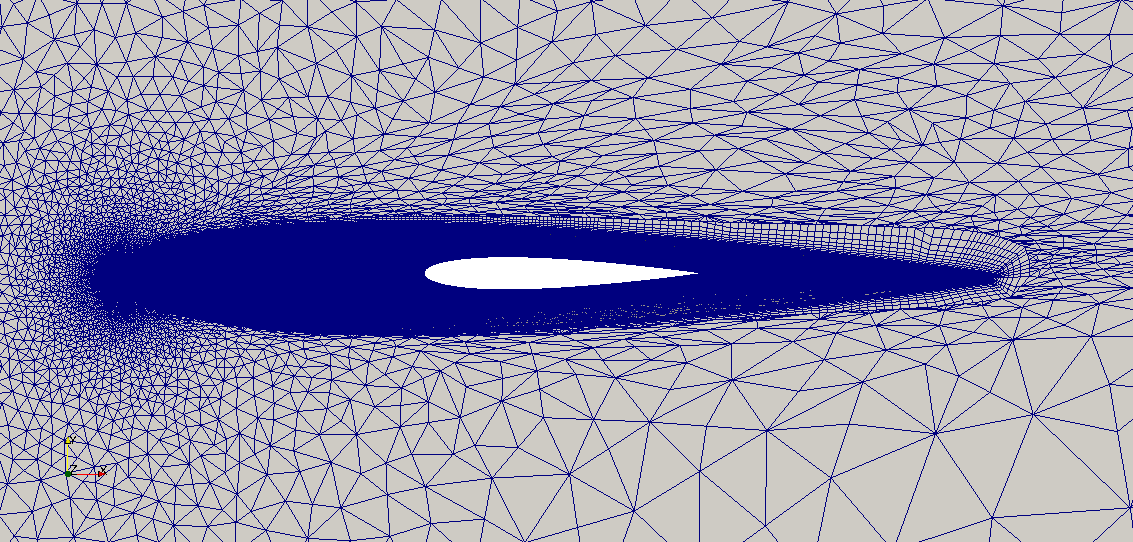}
     \label{f:NACA_10DegMesh}
 }
\end{center}
\vspace{-5mm}
 \caption{Initial and adapted meshes for NACA 0012}
 \label{f:NACA0012_AllMeshes}
\end{figure}

Figure~\ref{f:NACA0012_AllMeshes} shows the initial mesh and the second adapted meshes (LEV2) for each angle of attack (AoA). The initial mesh has a uniform coarse boundary layer mesh over the entire length of the airfoil. As the boundary layer mesh is adapted, for 0\degrees \hspace{0.1mm} AoA, the boundary layer heights on both sides of the airfoil are similar. Near the nose the boundary layer height remains low and increases as one moves towards the trailing edge. At 5\degrees \hspace{0.1mm}  AoA, the boundary layer on the suction side is thicker than the pressure side of the airfoil and this is clearly reflected in thickness adaptation of the mesh. As the angle of attack is further increased to 10\degrees \hspace{0.1mm}  AoA, the boundary layer on the suction side becomes thicker. This behavior as well is captured nicely by the adapted mesh. All the adapted meshes develop anisotropy in the streamwise direction in the unstructured part of the mesh. The stagnation point is refined more compared to other parts of the mesh. This is a result of using total scaled pressure Hessians instead of using velocity Hessians alone. 

%Figure~\ref{f:NACA0DegBLHeight} shows the comparison of our calculations for the boundary layer height with experiments conducted by Becker \cite{NACA}. Our calculations show good agreement with the experiments except at the start of the airfoil where the boundary layer is laminar. The calculations over predict the boundary layer height in the laminar region, which is not a bad thing because usually it is desirable to have the mesh boundary layer height greater than the physical boundary layer height.  
%
%\begin{figure}[h!]
%\begin{center}
% \includegraphics[width = 12cm]{NACA_BLHeight.png}
%\end{center}
% \caption{Comparison of boundary layer height calculations with experiments for NACA 0012 $0\degrees$  angle of attack}
% \label{f:NACA0DegBLHeight}
%\end{figure}

Table~\ref{t:NACA_yplus} shows the average $ y^+$ values of the first cells in the boundary layer for various meshes and different angles of attack. The average first cell height for the initial mesh is close to 3 for all angles of attack. As the thickness is adjusted with adaptation, this comes down to around 1.5 for the LEV1 mesh and to about 1.2-1.3 for the LEV2 mesh, which is close to the value of 1 that is targeted during adaptation. The behavior of adapting the first cell height towards $ y^+$ of 1 is consistent for all angles of attack. 

\begin{table}[h!]
\centering
\newcolumntype{A}{>{\centering\arraybackslash}m{3 cm}}
\newcolumntype{B}{>{\centering\arraybackslash}m{4 cm}}
\newcolumntype{D}{>{\centering\arraybackslash}m{2 cm}}
      \begin{tabular}{|A|B|A|A|}
      \hline
  	Angle of attack & 0\degrees & 5\degrees & 10\degrees  \\ \hline
	Initial mesh: LEV0 & 2.80 & 2.83 & 2.58  \\ \hline
	Adapted mesh: LEV1 & 1.51 & 1.49 & 1.58  \\ \hline
	Adapted mesh: LEV2 & 1.22 & 1.29 & 1.31 \\ \hline
      \end{tabular}
  \caption{Average first cell height in $y^+$ units}
  \label{t:NACA_yplus}
\end{table}

\begin{figure}[h!]
\begin{center}
\subfigure[0\degrees \hspace{0.1mm}  AoA]{
 \includegraphics[width = 16 cm]{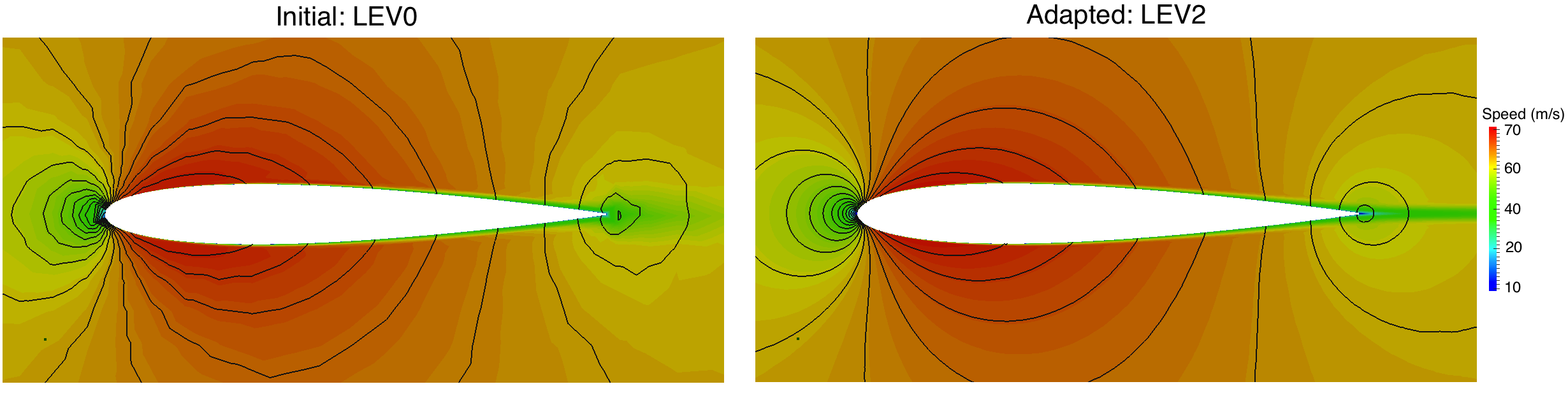}
  \label{f:NACA0DegVel}
 }
 \subfigure[5\degrees \hspace{0.1mm}  AoA]{
  \includegraphics[width = 16 cm]{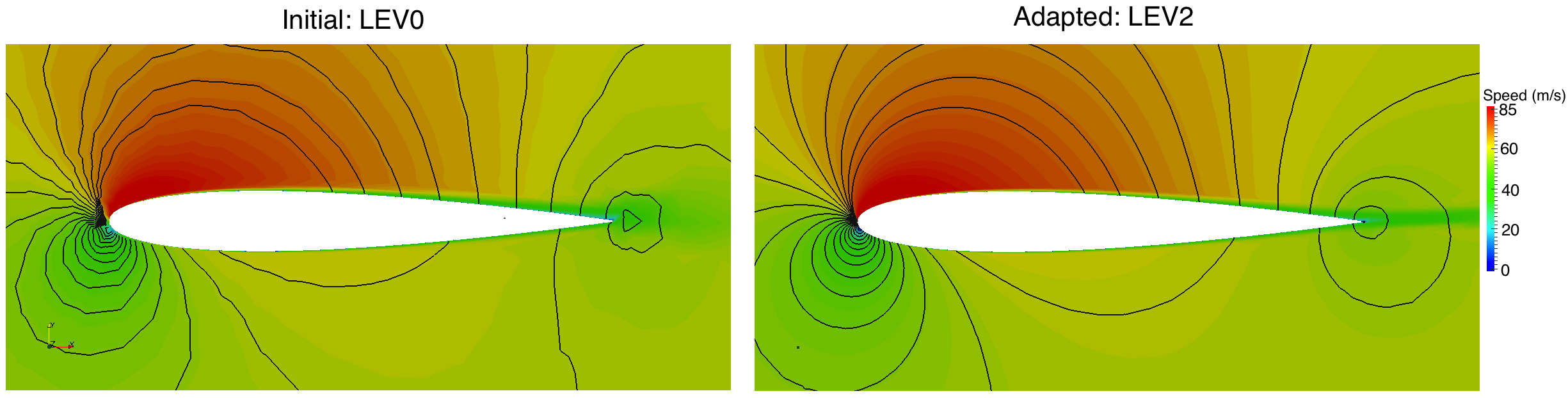}
   \label{f:NACA5DegVel}
 }
  \subfigure[10\degrees \hspace{0.1mm}  AoA]{
  \includegraphics[width = 16 cm]{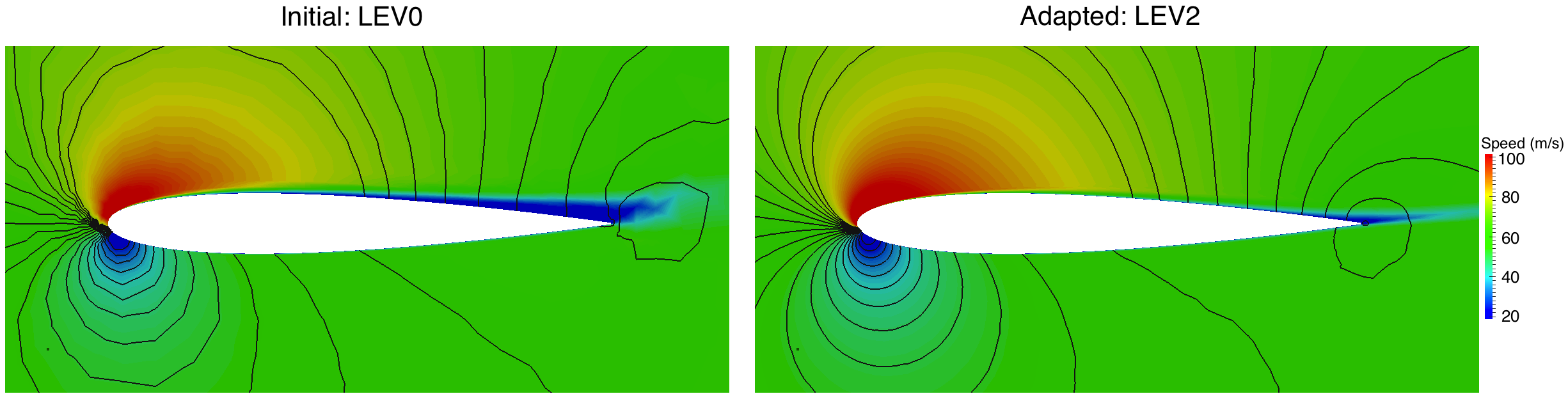}
   \label{f:NACA10DegVel}
 }
\end{center}
\label{f:NACA_AllVel}
 \caption{Slices of the initial LEV0 and adapted LEV2 meshes showing speed distributions with pressure contours (black lines)}
\end{figure}
Figure~\ref{f:NACA0DegVel} shows the speed distribution giving some indication of the height of the boundary layer and the pressure contours (black lines). The jagged pressure contours seen in the initial mesh become smoother with adaptivity. The adapted heights of the boundary layers on the upper and the lower surfaces of the wing are equal, as it should be for $0\degrees$ \hspace{0.1mm} angle of attack. The stagnation point and the wake receive more refinement than other areas. Anisotropic elements oriented with longer edges in the streamwise direction can be seen easily. 

As the angle of attack is increased to 5\degrees, the physical boundary layer on the suction surface has to face an adverse pressure gradient and starts getting thicker than the pressure surface. This behavior is correctly captured by our thickness adaptation strategies as already shown in Figure~\ref{f:NACA_5DegMesh} and the mesh boundary layer on the suction surface is adapted to be greater in height. The boundary layer on the pressure surface is comparatively smaller. This can be seen in Figure~\ref{f:NACA5DegVel}. Similar to 0\degrees \hspace{0.1mm}  the contours are captures better with adaptivity.

As the angle of attack is further increased to 10\degrees,  the boundary layer on the suction surface thickens much more than the one on the pressure surface. This behavior is captured correctly by our tools and is shown in Figure~\ref{f:NACA_10DegMesh}. The speed and pressure distribution are also enhanced with adaptivity. Since the boundary layer on the initial mesh is coarse, it predicts purious separation near the trailing edge. As the mesh is adapted in the boundary layer, it correctly predicts attached behavior. 

\begin{figure}[h!]
\begin{center}
\subfigure[AoA = 0\degrees]{
 \includegraphics[width = 8.0cm]{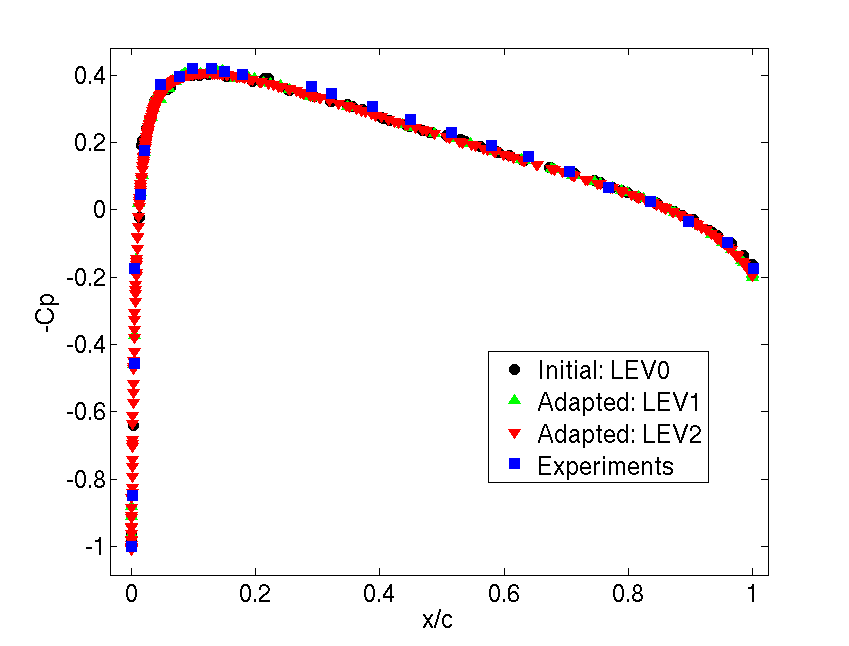}
 \label{f:NACA0DegCp}
}
\hspace{-10mm}
\subfigure[AoA = 10\degrees]{
  \includegraphics[width = 8.0cm]{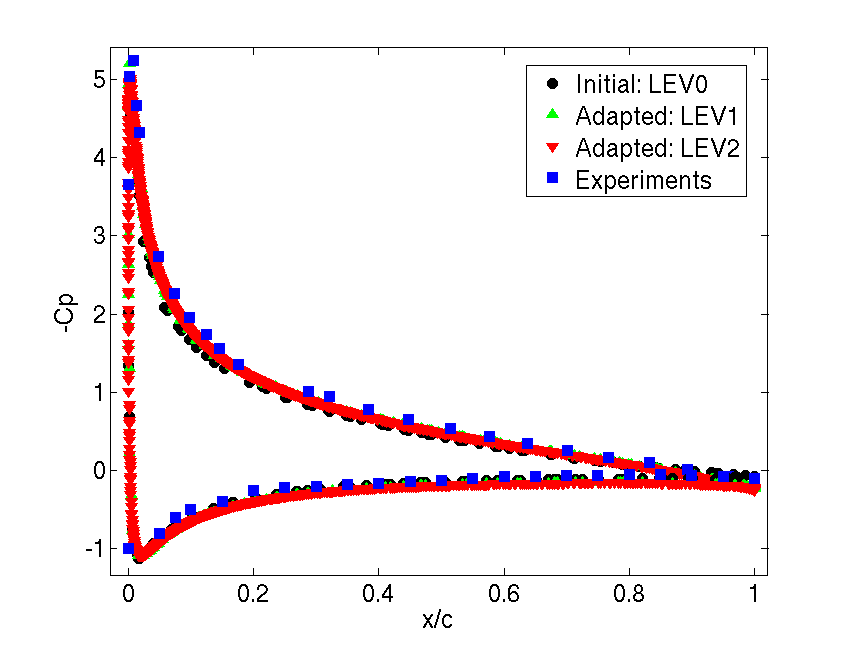}
  \label{f:NACA10DegCp}
}
\end{center}
 \caption{Coefficient of pressure plots}
 \label{f:NACACpAll}
\end{figure}

Figure~\ref{f:NACACpAll} shows the coefficients of pressure for 0\degrees \hspace{0.1mm} and 10\degrees \hspace{0.1mm}  angle of attack along with experiments. The experimental data for 0\degrees \hspace{0.1mm}  is from~\cite{Ladson}. For 10\degrees \hspace{0.1mm}  the pressure data on the upper surface is taken from~\cite{Gregory} and that on the lower surface is from~\cite{Ladson}. For 0\degrees, all meshes give a very good agreement with the experiments including the coarse initial mesh.  However, for 10\degrees, the suction pressure peak near the leading edge is better captured with the adapted meshes as compared to the initial mesh. On the suction side, the initial mesh predicts a much thicker boundary layer near the trailing edge which is close to incipient separation. The boundary layer profile close to the trailing edge shows negative X velocities for the initial mesh, which is not correct. This behavior is corrected in the adapted mesh due to better resolution in the boundary layer and no negative X velocities are detected. 

\begin{figure}[h!]
\begin{center}
 \includegraphics[width = 14 cm]{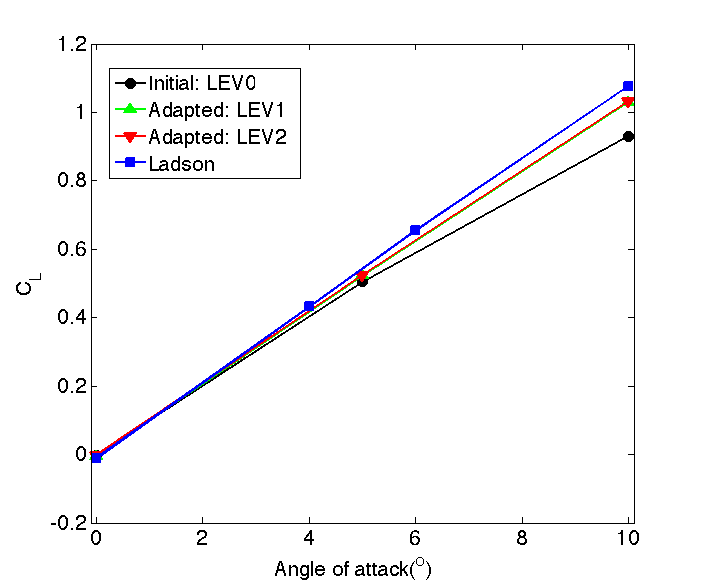}
\end{center}
 \caption{Coefficient of lift for NACA 0012 for different meshes in comparison to the experiment}
 \label{f:NACACL}
\end{figure}

To compare the results with experiments, the coefficient of lift curve is plotted in Figure~\ref{f:NACACL}. The experimental data is from Ladson~\cite{Ladson2}. At $0\degrees$ \hspace{0.1mm} angle of attack, all meshes show good agreement with the experiments. The initial mesh shows deteriorating $ C_L$ values at higher angle of attacks and adapted meshes show closer results to the experiments. The initial mesh fails to predict the correct $ C_L$ at $10\degrees$ \hspace{0.1mm} angle of attack but the adapted meshes show reasonable agreement. This wrong behavior of the initial mesh can be attributed to the false separation predicted on the suction side, near the trailing edge. As the boundary layer thickness is adjusted in the adapted mesh, this behavior is eliminated and we get better $ C_L$ values. The curves for the adapted LEV1 and LEV2 meshes lie on top of each other, indicating grid convergence. The small difference in the experimental and numerical values can be attributed to 3D effects arising from the finite width of the airfoil in the simulations. 

\section{CONCLUSIONS}

In this paper new physical indicators for boundary layer adaptation were described. The indicators take into account the turbulence modeling approach as well as the behavior of the physical boundary layer. Adaptivity in the unstructured region was driven by Hessians of the solution quantities. This approach was applied to a turbulent flat plate and three angles of attack for NACA 0012 airfoil. For the flat plate, adaptation significantly improved the boundary layer resolution which resulted in more accurate velocity profiles. The calculations of various quantities like $ C_f$ and $ \delta_{99}$ were compared to experiments to attest their accuracy. In case of the NACA 0012 airfoil,  adaptation gave more accurate results in terms of coefficient of lift. The changes in the mesh boundary layer thickness were highlighted in both test cases, which is a new development. 
\newpage
\bibliography{bibtex_database}
\bibliographystyle{plain}

\end{document}